\documentclass[aps,prb,twocolumn,groupedaddress]{revtex4-2}
\usepackage{graphicx}
\usepackage{epstopdf}
\usepackage{amsmath}
\begin{document}
\title{Current-induced phonon Hall effect}
\author{Kangtai Sun}
\thanks{\href{mailto:E0212212@u.nus.edu} {E0212212@u.nus.edu}}
\author{Zhibin Gao}
\author{Jian-Sheng Wang}
\affiliation{Department of Physics, National University of Singapore, Singapore 117551, Republic of Singapore}
\date{21 May 2020}
\revised{26 September 2020}

\begin{abstract}
Since the first experimental observation of the phonon Hall effect (PHE) in 2005, its physical origin and theoretical explanation have been extensively investigated.  While spin-orbit interactions are believed to play important roles under external magnetic fields, nonmagnetic effects are also possible.  Here, we propose a mechanism of PHE which is induced by electric current in a nonequilibrium system through electron-phonon interactions.  The influence of the drift electrons to the phonon degrees of freedom,  as a correction to the Born-Oppenheimer approximation, is represented by an antisymmetric matrix which has the same form as in a typical phonon Hall model.  We demonstrate the idea with a graphene-like hexagonal lattice having a finite phonon Hall conductivity under a driven electric current.

\end{abstract}
\maketitle

\section{Introduction}
The Hall effects, which have been widely studied in electronic systems, are
also observed and explained in recent years in phononic systems. The thermal
current could also be bent by a magnetic field  \cite{strohm2005phenomenological} through Raman-type spin-phonon interactions 
\cite{PhysRevLett.105.225901}.  As with the integer quantum Hall effect, the
phonon Hall effect can be related to the topological nature of the phonon
bands \cite{PhysRevLett.105.225901, PhysRevB.86.104305}.    More generally, parallel to the Hall effect in electron transport, it was proposed that, as long as there is a gauge potential playing a similar role as the vector potential in a magnetic field, there will be PHE \cite{PhysRevB.86.104305}. This net vector potential could come from the inner electron structure of an atomic system itself combined with an external magnetic field \cite{saito2019berry}, which has been observed in very recent experiment \cite{PhysRevLett.124.105901}, or other more complicated interactions like magnon-phonon interactions \cite{PhysRevLett.123.167202}. All of the present PHEs, either experimental or theoretical, need external \cite{agarwalla2011phonon, chen2020enhanced} or internal magnetic field to induce the observable phonon Hall conductivity. 

In 2010, L\"u $\emph{et~al.}$ \cite{doi:10.1021/nl904233u} applied an electric
current to a molecular junction and found that the current could break the
junction due to a nonconservative force, originated from a Berry phase.
This inspires us to think about what could happen if we apply an electric
current to a lattice system.  Having a current means we have broken the
time-reversal symmetry, which in some sense has the same effect as an applied
magnetic field.  For the Hall conductivity calculation, we follow the modern
method of Qin $\emph{et~al.}$ \cite{PhysRevB.86.104305}, which takes into
account the so-called energy magnetization contribution, while those of earlier
results of Wang and Zhang based on the Green-Kubo formula \cite{PhysRevB.80.012301,sheng2006theory,kagan2008anomalous} did not realize such a correction.  We compute the phonon Hall
conductivity and obtain an approximately linear dependence with the drift
velocity. 

The paper is organized as follows. In section II, we introduce a general theory
for the PHE and the principle of our current-induced PHE. In section III, we
demonstrate how we construct our lattice model. In section IV, we show our
numerical results and discuss their significance. In section V, we draw a brief
conclusion of our work. We also give an Appendix section which contains some
key details.

\section{Mechanism of Phonon Hall effect}

\subsection{Phonon Hall effect under non-zero vector potential}

What is the most general form of a Hamiltonian for phonons that can result in a
Hall effect?  Let us consider a very general system described by $2N$ Hermitian
variables $y_j$, $j=1,2,\cdots, 2N$, for a system of $N$ degrees of freedom.
In column vector notation, we denote this by $y$, where x components come first, then followed by y components for each degree of freedom.  We assume that the Hamiltonian
takes a quadratic form of $\hat H = \frac{1}{2} y^T H y$, here we assume $H$ is
real and symmetric, superscript $T$ is the matrix transpose.  The operators $y_j$ are completely characterized by their
commutation relations, $[y_j,y_{j^{\prime}}] = i\hbar J_{jj^{\prime}}$.   We assume that $J_{jj^{\prime}}$
is a $c$-number.  Since $y$ is Hermitian, we can show that the matrix $J$ is
real and antisymmetric.  The Heisenberg equation of motion is simply 
\begin{equation} 
{dy \over dt}=  J H y.
\end{equation} 

Two common choices of $y$ appear in the literature, that of Zhang $\emph{et~al.}$ use conjugate
pairs of displacement coordinates $u$ and momenta $p$, while Qin $\emph{et~al.}$ use
the displacements $u$ and velocities $v=du/dt = p - Au$.  Here in this paper, we
follow Qin's convention.  Then the matrix $J$ takes the following form: 
\begin{equation} 
J =
\begin{pmatrix} 0 &I\\ -I &-2A \end{pmatrix}, \quad{\rm with}\quad
y=\begin{pmatrix} u \\ v \end{pmatrix},  
\end{equation} 
here the matrix $A$ is antisymmetric. 

The effect of the Berry phase was long known in coupled electron-nuclear systems \cite{mead1992geometric}, but usually, this extra term is neglected in a
Born-Oppenheimer approximation.  When this term is taken back, the
Hamiltonian of the nuclei or phonons in a solid is given by Mead and Truhlar 
\cite{doi:10.1063/1.437734}:
\begin{equation}
\hat{H}=\sum\limits_{lj}\frac{(-i\hbar\boldsymbol{\nabla}_{lj}-\boldsymbol{A}(\boldsymbol{R})_{lj})^2}{2M_j}+U(\boldsymbol{R}),
\end{equation} 
where $\boldsymbol{R}_{lj}$ is the nucleus position vector of atom $j$ with
mass $M_j$ in the unit cell $l$, $U(\boldsymbol{R})$ is the potential on the 
nuclei. Here the vector potential $\boldsymbol{A}$ comes from the electron
Berry phases but can also be the effect of other interactions such Raman-type
spin-phonon interaction, external magnetic fields \cite{PhysRevLett.105.225901},
or spin-orbit interaction within electronic structure 
\cite{PhysRevB.86.104305}. Through out this paper, index $j$ for bold symbol stands for atom sites, for unbold symbol, $j$ also includes Cartesian components. In a periodic lattice system with a harmonic
approximation, we can transform the system into the reciprocal space, and use a
combined coordinate and velocity variable  $y_{\boldsymbol{q}}$ so that
$\hat{H}=\frac{1}{2}\sum_{\boldsymbol{q}}y_{\boldsymbol{q}}^{\dagger}H(\boldsymbol{q})y_{\boldsymbol{q}}$.
Here $\boldsymbol{q}$ is the wavevector sampling over the first Brillouin
zone.  Note that $y_{\boldsymbol{q}}$ is not a Hermitian operator; it is a
vector of smaller dimension varying over twice the degrees of freedom per unit
cell for each $\boldsymbol{q}$.  Elements of the $H(\boldsymbol{q})$ matrix are
determined by $y_{\boldsymbol{q}}$.  The commutation relation in
$\boldsymbol{q}$ space is \cite{PhysRevLett.123.167202}
\begin{equation}
[y_{j\boldsymbol{q}},y_{j^{\prime}\boldsymbol{q}^{\prime}}^\dagger]=i\hbar J_{jj^{\prime}}(\boldsymbol{q})\delta_{\boldsymbol{q}\boldsymbol{q}^{\prime}}.
\end{equation}

Next by assuming $y_{\boldsymbol{q}}=\psi_{\boldsymbol{q}}e^{-i\omega t}$, the
corresponding eigensystem of the equation of motion will be
\begin{equation}
iJ(\boldsymbol{q})H(\boldsymbol{q})\psi_{\boldsymbol{q}}\equiv H_{\rm eff}\psi_{\boldsymbol{q}}
= \omega\psi_{\boldsymbol{q}}.
\end{equation}
Since the effective Hamiltonian is non-Hermitian, the left eigenvector is not related by Hermitian conjugate to the right eigenvector.  We can choose the left eigenvector as  
$\bar{\psi}_{\boldsymbol{q}} = \psi_{\boldsymbol{q}}^{\dagger}H(\boldsymbol{q})$.  The normalization condition is then 
$\psi_{\boldsymbol{q}}^{\dagger}H(\boldsymbol{q})\psi_{\boldsymbol{q}}\equiv\bar{\psi}_{\boldsymbol{q}}\psi_{\boldsymbol{q}}=1$.
This eigen equation is general to any possible source of the non-zero vector
potential. For example, we can choose
$y_{\boldsymbol{q}}=(\boldsymbol{u}_{\boldsymbol{q}},\boldsymbol{v}_{\boldsymbol{q}})^T$
where $\boldsymbol{v}_{\boldsymbol{q}}=\dot{\boldsymbol{u}}_{\boldsymbol{q}},
\boldsymbol{u}_{j\boldsymbol{q}}=\sqrt{M_j/N}\sum_l\boldsymbol{x}_{lj}e^{-i\boldsymbol{q}\cdot\boldsymbol{R}_{l}^0}$
with $\boldsymbol{R}_{l}^0$ being the real space lattice vector, 
$\boldsymbol{x}_{lj}$ being the deviation from equilibrium positions of atom $j$ in cell $l$.   $N$ is the total number of unit cells. We write   
$\boldsymbol{u}_{\boldsymbol{q}}$ without the index $j$ as a column vector consisting of the degrees in a unit cell.   Once we have obtained the eigenvalues and associated eigenvectors of the effective Hamiltonian, we can calculate its Berry curvature and phonon Hall conductivity using the formulas given by Qin $\emph{et~al.}$ \cite{PhysRevB.86.104305},
\begin{equation}
\boldsymbol{\Omega}_{\boldsymbol{q}i} =-\text{Im}\Big[\frac{\partial\bar{\psi}_{\boldsymbol{q}i}}{\partial\boldsymbol{q}}\times\frac{\partial\psi_{\boldsymbol{q}i}}{\partial\boldsymbol{q}}\Big],
\end{equation}
and \cite{zhang2016berry}
\begin{equation}\label{eq:kappa}
\kappa_{xy}=-\frac{1}{2T}\int_{-\infty}^\infty d\epsilon\epsilon^2\sigma_{xy}(\epsilon)\frac{dn(\epsilon)}{d\epsilon},
\end{equation}
where
\begin{equation}
\sigma_{xy}(\epsilon)=-\frac{1}{V\hbar}\sum\limits_{\hbar\omega_{\boldsymbol{q}i}\le\epsilon}\Omega_{\boldsymbol{q}i}^z,
\end{equation}
$n(\epsilon) = 1/(e^{\epsilon/(k_BT)} - 1)$ is the Bose function at temperature
$T$,  and $k_B$ the Boltzmann constant. In the above summation over mode $\boldsymbol{q}i$, all modes with both positive and negative frequencies, are included.  Since we are dealing with a two-dimensional sheet, the volume $V$ is an ill-defined concept.  We use $V = L^2 a$, the area times the thickness, choosing $a$ somewhat arbitrarily to match the units of W/(mK) of the usual
three-dimensional thermal conductivity. When estimating the phonon Hall conductivity $\kappa_{xy}$, we assume the thickness of the sample is the 
same as the bond length $a=1.42$ \AA\ of a graphene lattice.

\subsection{Current-induced non-zero vector potential}

\begin{figure}[h!]
\centering
\includegraphics[scale=0.6]{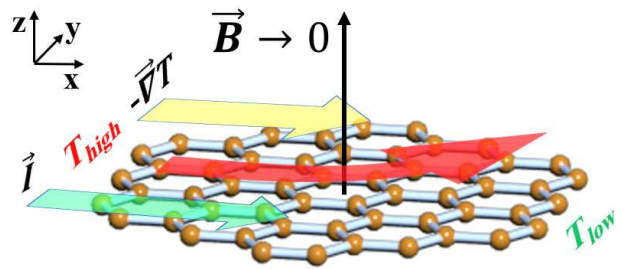}
\caption{The schematic setup to detect current-induced phonon Hall effect. Electric current and temperature gradient are needed which are parallel to each other. A very small magnetic field, which is about ~$10^{-5}$ tesla, is to perturb the system and distinguish the direction of the phonon Hall current.\label{fig:Modelsetting}}
\end{figure}

L\"u $\emph{et~al.}$ \cite{doi:10.1021/nl904233u} theoretically studied the
effect of electric current on a molecular bridge connecting two metallic
electrodes. They found a new mechanism, which involves Berry phase, that can
lead to a breakdown of the bridge by a ``run away'' mode. Their discovery
inspired us to ask if we introduce electric current into a lattice system,
$\emph{e.g.}$, the honeycomb lattice, is there a phonon Hall effect? The ``run away'' mode means the amplitude of oscillation including those perpendicular to the molecular bridge will grow in time, therefore if we extend it to a 2D lattice, this ``run away'' mode induced by electric current may result in a phonon Hall current. Figure \ref{fig:Modelsetting} provides a possible setup on a honeycomb lattice for this current-induced phonon Hall effect. 

For convenience, we use the renormalized coordinate $\boldsymbol{u}_{lj}\equiv\sqrt{M_j}\boldsymbol{x}_{lj}$ to
denote the nucleus displacement in real space.  Electrons in a metal or a
semi-conductor carrying electric current can interact with the lattice phonons
through the electron-phonon interaction (EPI). In the NEGF formalism, EPI
effect is included as a self-energy term in the phonon retarded Green's function
\cite{RevModPhys.89.015003},
\begin{equation}
\text{D}(\omega,\boldsymbol{q})=\bigl[\omega^2 I -\tilde{K}_{\boldsymbol{q}}-\Pi(\omega\!=\!0)-\Pi^{\text{NA}}_{\boldsymbol{q}}(\omega)\bigr]^{-1},
\end{equation}
where $I$ is the identity matrix in site space of a unit cell, $\tilde{K}_{\boldsymbol{q}}$ is the dynamic matrix. $\Pi(\omega\!=\!0)$
is the second term in the equation below. We subtract it off so that the leading
contribution is proportional to the frequency $\omega$ in the so-called
non-adiabatic self-energy due to electrons: 
\begin{equation}
\begin{aligned}
\Pi^{\text{NA}}_{\boldsymbol{q}jj^{\prime}}&(\omega)={1 \over N}\sum\limits_{mn}\sum\limits_{\boldsymbol{k}} g^*_{mnj}(\boldsymbol{k},\boldsymbol{q})g_{mnj^{\prime}}(\boldsymbol{k},\boldsymbol{q})\\
&\times\Big[\frac{f_{m\boldsymbol{k+q}}-f_{n\boldsymbol{k}}}{\varepsilon_{m\boldsymbol{k+q}}-\varepsilon_{n\boldsymbol{k}}-\hbar\omega-i\eta}-\frac{f_{m\boldsymbol{k+q}}-f_{n\boldsymbol{k}}}{\varepsilon_{m\boldsymbol{k+q}}-\varepsilon_{n\boldsymbol{k}}}\Big],
\end{aligned}
\end{equation}
where $f$ is the Fermi function, $g$ is the converted EPI matrix falling in electron mode space and phonon reciprocal space, $\boldsymbol{k}$ and
$\boldsymbol{q}$ are wave vectors of electrons and phonons respectively,
$\varepsilon_{n\boldsymbol{k}}$ is the electron dispersion relation, the
subscripts $m$ and $n$ indicate the electron bands, and the subscripts $j$
and $j'$ denote the atomic labels in a unit cell including both atom sites and Cartesian directions. The summation is over the first Brillouin zone of the electrons. A small positive $\eta$ attributes the electrons with a finite life time. The self energy can be computed from a first-principle package.

Alternatively, the movement of the ions can also be described semi-classically
by an equation of motion taking into account the effect of the electrons.  In
real space under a Markov approximation, it takes the form
\cite{doi:10.1063/1.4917017},
\begin{equation}
\ddot{\boldsymbol{u}}=-K\boldsymbol{u}-2A\dot{\boldsymbol{u}},
\end{equation}
where $K$ is the spring constant matrix in real space corresponding to the dynamic matrix 
$\tilde{K}_{\boldsymbol{q}}$ in reciprocal space, and $A$ can be regarded as
the matrix representation of the vector potential induced by EPI which is
antisymmetric.  Therefore, the phonon Green's function is:
\begin{equation}
\text{D}(\omega, \boldsymbol{q})=\bigl[\omega^2 I -\tilde{K}_{\boldsymbol{q}}+2i\omega \tilde{A}_{\boldsymbol{q}}\bigr]^{-1}.
\end{equation}
Comparing the two expressions, if we ignore the higher order terms of $\omega$
in $\Pi^{\text{NA}}(\omega)$, and note that $\tilde{A}_{\boldsymbol{q}}$ is
anti-Hermitian (the anti-Hermitian part of $\Pi^{\text{NA}}(\omega)$ is the source of
dissipative Joule heating, which we will ignore.), we can conclude that:
\begin{equation}
\tilde{A}_{\boldsymbol{q}}=\lim_{\omega\rightarrow0}\frac{\Pi^{\text{NA}}(\omega)+(\Pi^{\text{NA}})^{\dagger}(\omega)}{-4i\omega}.
\end{equation}
The Markov approximation adopted here is well justified as the electrons move
on a much faster time scale than that of the nuclear degrees of freedom.  In
terms of the energy scale, an electron has typical energy of order eV, while
phonon $\hbar \omega$ is of the order 100\,meV or less. So keeping the leading $\omega$
dependence only on self-energy is a good approximation.  We can trace back to an
effective Hamiltonian for phonons with the electrons taken into account through a
non-dissipative term as
\begin{equation}
\hat{H}=\frac{1}{2}(p-Au)^2+\frac{1}{2}u^TKu,
\end{equation}
and the corresponding eigen equation is
\begin{equation}
\begin{aligned}
\omega\psi_{\boldsymbol{q}}&=i\begin{pmatrix} 0 &I\\ -I &-2\tilde{A}_{\boldsymbol{q}} \end{pmatrix}\begin{pmatrix} \tilde{K}_{\boldsymbol{q}} &0\\ 0 &I \end{pmatrix}\psi_{\boldsymbol{q}}\\
&=\begin{pmatrix}
0 &iI\\ -i\tilde{K}_{\boldsymbol{q}} &-i2\tilde{A}_{\boldsymbol{q}} \end{pmatrix}\psi_{\boldsymbol{q}}.
\end{aligned}
\end{equation}
Here we choose $y_{\boldsymbol{q}}=(\boldsymbol{u}_{\boldsymbol{q}},
\boldsymbol{v}_{\boldsymbol{q}})^T$, and $\boldsymbol{v}_{\boldsymbol{q}} =
\boldsymbol{p}_{\boldsymbol{q}} -\tilde
A(\boldsymbol{q})\boldsymbol{u}_{\boldsymbol{q}}$ as before.

\section{Model implementation on a graphene-like lattice}

\subsection{Hamiltonians and self-energy}

Graphene has been widely studied and it has remarkably high electron mobility,
therefore we choose a graphene-like lattice to implement our settings. We use a
standard spinless tight-binding model for the electrons:
\begin{equation}
\hat{H}_e=-t\sum\limits_{l\delta}\big[c_{A,l}^{\dagger}c_{B,l+\delta}+c_{B,l}^{\dagger}c_{A,l+\delta}\big],
\end{equation}
where $t=2.8\,\text{eV}$ is the hopping parameter. $A$ and $B$ indicate the two
sublattices, and $l$ runs over the Bravais lattice sites and $\delta$ runs over
the displacements of the three nearest neighbors of a given site.  Zhang
$\emph{et~al.} $\cite{PhysRevLett.105.225901} have proposed a simple phonon
model for a graphene-like lattice in which the coupling matrix is diagonal when
the bond orientation is in the $x$ direction between two atoms,
\begin{equation}
K_x=\begin{pmatrix} K_L &0\\ 0 &K_T \end{pmatrix},
\end{equation}
where $K_L=0.144 \text{ eV}/(\text{u\AA}^2)$ is the longitudinal spring
constant and $K_T=K_L/4$ is the transverse spring constant.   Other
orientations can be obtained by rotations.  The dynamic matrix is given by
\begin{equation}
\tilde K_{\boldsymbol{q}} = \sum\limits_{l^{\prime}}K_{ll^{\prime}}e^{i(\boldsymbol{R}_{l^{\prime}}^0-\boldsymbol{R}_{l}^0)\cdot\boldsymbol{q}},
\end{equation}
where $K_{ll^{\prime}}$ is the submatrix between unit cell $l$ and $l^{\prime}$ in the full $K$. In this model, we have ignored the $z$ mode and consider only the in-plane
motion.  The reason is that the motion in the direction perpendicular to the plane
couples quadratically to the electron degrees of freedom, and this is a high order
effect to the electron-phonon interaction.

For the electron-phonon interaction, we take a Su-Schrieffer-Heeger-like model,
as used in a previous work by Jiang and Wang \cite{doi:10.1063/1.3671069},
\begin{equation}
\begin{aligned}
\hat{H}_{\rm epi}=J_1\sum\limits_{l\delta}&\big[c_{A,l}^{\dagger}c_{B,l+\delta}+c_{B,l+\delta}^{\dagger}c_{A,l}\big]\\
&\times[(\boldsymbol{u}_{B,l+\delta}-\boldsymbol{u}_{A,l})\cdot\boldsymbol{\hat{e}}_{l,\delta}],
\end{aligned}
\end{equation}
where $J_1=-6.0\,\text{eV/\AA}$ and $\boldsymbol{\hat{e}}_{l,\delta}$ is the
direction between two nearest atoms.  The $g$ matrix
is given by
\begin{equation}
g_{mnj}(\boldsymbol{k}, \boldsymbol{q}) = \sum\limits_{m^{\prime}n^{\prime}}S^{\dagger}_{mm^{\prime}}(\boldsymbol{k}+\boldsymbol{q})\Xi^j_{m^{\prime}n^{\prime}}(\boldsymbol{k}, \boldsymbol{q})S_{n^{\prime}n}(\boldsymbol{k}),
\end{equation}
where $j = \{\rm Ax, Ay, Bx, By\}$,
\begin{equation}
S(\boldsymbol{k})={1 \over \sqrt{2}} \begin{pmatrix} 1 &e^{i\phi(\boldsymbol{k})} \\ -e^{-i\phi(\boldsymbol{k})} &1 \end{pmatrix},
\end{equation}
with $e^{i\phi(\boldsymbol{k})} = f(\boldsymbol{k})/|f(\boldsymbol{k})|$, $f(\boldsymbol{k})=e^{-ik_xa}+e^{i(k_xa/2+\sqrt{3}k_ya/2)}+e^{i(k_xa/2-\sqrt{3}k_ya/2)}$, and $\Xi^j_{m^{\prime}n^{\prime}}(\boldsymbol{k}, \boldsymbol{q})$ is the reciprocal EPI matrix corresponding to $\hat{H}_{\rm epi}$. The expression is given in Appendix A. 

In this work, we focus on the EPI for $\boldsymbol{k}$ points near the Dirac points
of the electrons and $\boldsymbol{q}$ near the $\Gamma$ point of the phonons, for we
find that they are dominant in determining the final phonon Hall conductivity.
It seems that we have prepared all the ingredients to calculate
$\tilde A_{\boldsymbol{q}}$.  However, there is a problem that when we apply an
electric current to this graphene-like two-dimensional surface, assuming the
drift velocity $v_1$ of current is along the $x$ direction, it is in a
nonequilibrium state, therefore we cannot just substitute the Fermi function
into the formula.  To solve this problem, we use a single-mode relaxation
approximation \cite{ziman2001electrons} so that:
\begin{equation}
f=f^0-\frac{\partial f^0}{\partial\varepsilon}\Phi\approx f^0(\varepsilon-\Phi),
\end{equation}
where $f^0=[e^{(\varepsilon-\mu)/k_BT}+1]^{-1}$ with $\mu$ being the chemical potential of electron, and $\Phi\equiv\Phi_{n\boldsymbol{k}}$ is mode dependent:
\begin{equation}
\Phi_{n\boldsymbol{k}}=-eE\tau_{nk}\frac{\partial\varepsilon_{n\boldsymbol{k}}}{\partial\hbar k_x},
\end{equation}
where $E$ is the applied electric field, $\tau_{nk}$ is the relaxation time
which is only related to the magnitude of the wave vector. In practice, since we don't know the relaxation time, we combine it with the electric field and
replace them with the drift velocity $v_1$, for graphene-like lattice
\cite{peng2018current}:
\begin{equation}
\Phi_{n\boldsymbol{k}}=v_1\,\text{Re}\Big[z^*\frac{\partial z}{\partial k_x}\Big]/(\hbar v_F^2),
\end{equation}
where $v_F=3at/(2\hbar)$ is the Fermi velocity, $a=1.42\text{ \AA}$ is the
distance between atoms, and
$z=-tf(\boldsymbol{k})$.
By requiring this correction to the Fermi function, the self-energy can be
numerically calculated, and thereafter, the $\tilde{A}_{\boldsymbol{q}}$ matrix.

\subsection{The Berry curvature - is it unique?}

\begin{figure}[h!]
\centering
\includegraphics[scale=0.5]{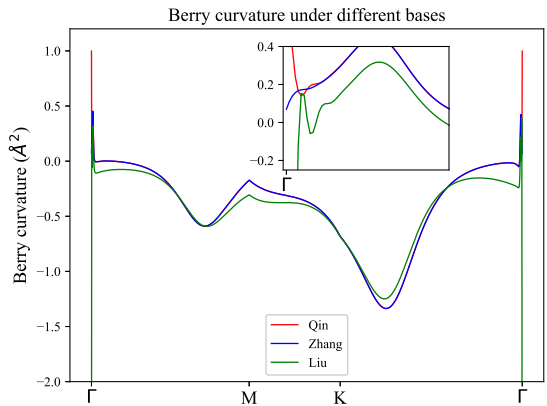}
\caption{The Berry curvatures along the high-symmetry path under three
different bases \cite{PhysRevLett.105.225901,PhysRevB.86.104305,
liu2017model}. Although they do not differ so much from each other, they are
indeed different. The parameter set is chosen to be: $\boldsymbol{b} \cdot \boldsymbol{q} = (1000 \,{\rm \AA},1 \,{\rm \AA}) \cdot \boldsymbol{q}, c=0.1$ rad/ps, and $\Lambda$  is a constant antisymmetric matrix with upper triangular elements, lower triangular elements and diagonal elements being 1.0, -1.0, and 0 rad/ps respectively. \label{fig:BCcompare}}
\end{figure}

As we have discussed in the previous section, the choice of
$y_{\boldsymbol{q}}$ is not unique -- at least three different choices exist in
the literature.  Zhang $\emph{et~al.}$ choose
$y_{\boldsymbol{q}}=(u_{\boldsymbol{q}}, p_{\boldsymbol{q}})$, Qin
$\emph{et~al.}$ choose $y_{\boldsymbol{q}}=(u_{\boldsymbol{q}},
v_{\boldsymbol{q}})$, Liu $\emph{et~al.}$ choose
$y_{\boldsymbol{q}}=(\tilde{K}_{\boldsymbol{q}}^{-\frac{1}{2}}u_{\boldsymbol{q}},
v_{\boldsymbol{q}})$ \cite{PhysRevLett.105.225901,PhysRevB.86.104305,
liu2017model}. The difference between Zhang's and Qin's choices is like the
difference between Lagrangian mechanics and Hamiltonian mechanics, therefore
they are more or less equivalent. The special choice of Liu results in a
Hermitian effective Hamiltonian, which implies immediately the eigenfrequencies are all real.  When the vector potential term can be separated from
the usual potential energy term as in our case, these three bases are related
by similarity transformations explicitly.  However, this kind of variable
transformations is not gauge invariant.  Therefore, generally, if
$\tilde{A}_{\boldsymbol{q}}$ is not a constant matrix, they will result in
different Berry curvatures. The question then arises as which one should be
used to compute the phonon Hall conductivity?  To illustrate and confirm that
there is indeed a difference, we choose a smooth $\tilde{A}_{\boldsymbol{q}} = (\Lambda+i|\Lambda|)*(\boldsymbol{b} \cdot \boldsymbol{q} + c)$ matrix, where $\Lambda$ is a constant $4 \times 4$ antisymmetric matrix, $|\Lambda|$ takes the absolute value of each element in $\Lambda$, $\boldsymbol{b}$ is a constant vector parameter, and $c$ is another constant parameter. In principle, these three bases should result in different Berry curvatures, but in practice, the differences are small, especially between Zhang's and Qin's choices, therefore we choose such a highly anisotropic case. We plot the corresponding Berry curvatures of the three bases along a high-symmetry path of the graphene-like lattice in Fig.\ref{fig:BCcompare}. We see that there are sharp peaks at the $\Gamma$ point.  However, the signs of the peaks are opposite for Liu et al.\@ definition to that of Zhang and Qin $\emph{et~al.}$.  Away from the $\Gamma$ point, the values tend to be close among the three. In conclusion, since only Qin $\emph{et~al.}$ derived the correct formula for the phonon Hall conductivity with their definition of the Berry curvature, which considers an energy magnetization contribution to Hall conductivity \cite{PhysRevB.86.104305} while Zhang $\emph{et~al.}$ did not, we prefer to follow Qin's definition. It is natural that if we use other choices, we will obtain different formulas for phonon Hall conductivity. 

\begin{figure}[h!]
\centering
\includegraphics[scale=0.5]{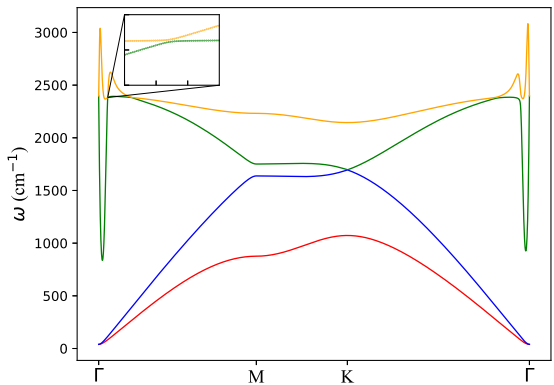}
\caption{The dispersion relation of positive branches along high-symmetry path
$\Gamma-M-K-\Gamma$ with $v_1=1.0\times10^4$ m/s, $T=300\,$K, $\mu = 0.1\,$eV. A small onsite potential $V_{\rm
onsite}=1.0\times10^{-3}K_L$ and a nearly 0 magnetic field measured by
effective parameter $h=1.0\times10^{-9}$ rad/ps are employed to perturb the
system. The inset shows one of the anti-crossing points. Note that the out-of-plane ZA mode is not considered here. \label{fig:dispersion}}
\end{figure}

\section{Numerical results and discussion}
In order to have a well-defined topological structure, we need to perturb our
system to open tiny gaps at $\Gamma$ and $K$ points, as the Berry curvature
becomes ill-defined when the bands are degenerate.  This goal is achieved by
adding a small onsite potential term to the phonon dynamic matrix and a nearly
zero magnetic field which goes into the Hamiltonian through Raman-type
spin-phonon interaction \cite{PhysRevLett.105.225901}.  The effect of the magnetic field is described by a constant antisymmetric matrix $A_h$:
\begin{equation}
A_h = \begin{pmatrix} B_h &0 \\ 0 &B_h \end{pmatrix},
B_h = \begin{pmatrix} 0 &h \\ -h &0 \end{pmatrix},
\end{equation} 
where $h$ is an effective parameter representing magnetic field with units rad/ps (1 rad/ps $\approx$ 33.3 cm$^{-1}$). Adding this matrix to our previous $\tilde{A}_{\boldsymbol{q}}$ will introduce magnetic field into our system. When we calculate $\tilde{A}_{\boldsymbol{q}}$, a $400\times400$ $\boldsymbol{k}$ grids is used and the parameter $\eta$ is set to be about ~0.2 eV. We note that as a function of a constant magnetic field $h$, the Berry curvatures and the Chern numbers are odd functions of $h$ and experience a discontinuity at $h=0$, thus ill-defined at $h=0$.   Our results presented below thus should be considered as the limit when $h \to 0^+$ and $V_{\rm onsite} \to 0^+$.   This is physical since we can always apply a small magnetic field and put the system on a substrate, thereby acquiring an onsite interaction. There is one more important thing to note that inside the formula of $\tilde{A}_{\boldsymbol{q}}$, since we only focus on $\boldsymbol{q}$ points near $\Gamma$ point, there is a hidden $\delta$ function behavior when temperature is low. This $\delta$ function originates from the difference of the intra-band Fermi functions in the numerator of $\tilde{A}_{\boldsymbol{q}}$ if we take a Taylor expansion of $\boldsymbol{q} $ near $\Gamma$ point at low temperature. To handle this $\delta$ function numerically, we should compute in a very dense $\boldsymbol{k}$ grids which requires a lot of computation power. However, we can also broaden this $\delta$ function by tuning the electron parameter $\beta=1/k_BT$. Through computation, we find that the differences of EPI at low temperature range, e.g., below 300 K or even below 500 K, are very small, therefore, when we calculate $\tilde{A}_{\boldsymbol{q}}$ at low temperature, we can make an approximation to fix the broadening parameter to be the value at higher temperature like 300 K or 500 K.

Figure \ref{fig:dispersion} shows the positive part of the dispersion relation
of our current-induced system, from which we can see that the two acoustic
branches are very close to the pure phonon system without the drift current, while the
two optical branches get modified drastically. This behavior is easy to
understand if we review the EPI form of our model. The strength of EPI in our model
is proportional to the relative displacement of atoms, therefore the optical
modes, in which atoms move relatively, are equipped with stronger EPI
than acoustic ones. It deserves notice that there are several anti-crossing
points in the dispersion relations.  These points will possess much larger Berry
curvature, therefore they are dominant in determining the topological
properties of the system. Points in acoustic branches near $\Gamma$ point and anti-crossing points near K points also have large Berry curvatures. However, these pairs of Berry curvatures should cancel each other for they are similar to pure phonon system where there are no PHE.

\begin{figure}[h!]
\centering
\includegraphics[scale=0.3]{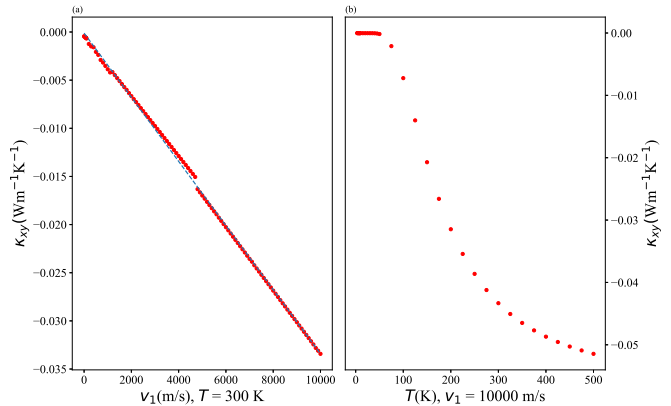}
\caption{(a) Phonon Hall conductivity $\kappa_{xy}$ versus drift velocity $v_1$
at a temperature $T=300\,$K. The broadening parameter is $\beta=1/(k_B\times300{\rm K})$. (b) Phonon Hall conductivity $\kappa_{xy}$ versus temperature at $v_1=10000\,$m/s. The broadening parameter is set to be $\beta=1/(k_B\times500{\rm K})$. These two plots share the same set of parameters of temperature, chemical potential, onsite potential and nearly 0 magnetic field as Fig.\ref{fig:dispersion}. \label{fig:Kxy-v1-T}}
\end{figure}

Figure \ref{fig:Kxy-v1-T}(a) demonstrates the relationship between
$\kappa_{xy}$ and the drift velocity $v_1$. $\kappa_{xy}$ is roughly linear dependent on $v_1$ for our picked velocity sequence. When $v_1$ is gradually close to the Fermi velocity of this graphene-like lattice system, our theory and approximation on EPI will gradually break down. The Chern numbers of positive branches are $C^1=1, C^2=C^3=0, C^4=-1$, where larger indices are associated with higher frequencies. In our range of the drift velocity, there is no jump among Chern numbers, which seems kind of trivial. The discontinuities are due to numerical errors for the Chern numbers do not change, which means the dispersion relation of the system has the same pattern. Figure \ref{fig:Kxy-v1-T}(b) shows temperature dependence of $\kappa_{xy}$. When the temperature is very small, PHE tends to disappear, and in our temperature range, the absolute value of the phonon Hall conductivity gradually increases as temperature is increasing, but we can not conclude what the exact relationship between $\kappa_{xy}$ and temperature is. In our calculation, numerical errors mainly come from the calculation of $\tilde{A}(\boldsymbol{q})$ and cubic interpolation to obtain its values with denser grids, which is 2000$\times$2000.

The order of magnitude of our current-induced $\kappa_{xy}$ is one order smaller than the case with the magnetic field parameter $h$ being several rad/ps. It is instructive to compare the magnitude of the Hall conductivity to the universal conductance quantum which is $G_0 = T (\pi k_B)^2/(3 h)$, when 
converted into the same units of conductivity, $G_0/a$, at 300\,K, we find it is 
about 2\,W/(mK). Our result is about 1/100-th of the conductance quantum.  
Since $\kappa_{xy}$ with our model is only about one order smaller than a pure magnetic field experimental results \cite{strohm2005phenomenological},  it should be still observable experimentally in principle. 

\begin{figure}[h!]
\centering
\includegraphics[scale=0.5]{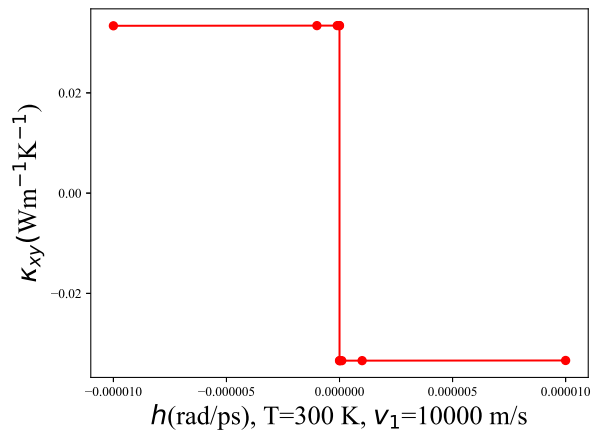}
\caption{Phonon Hall conductivity $\kappa_{xy}$ versus magnetic field parameter $h$. We can see $\kappa_{xy}$ changes sign as $h$ changes sign and there is a discontinuity when $h$ crosses 0. \label{fig:Kxy-h-Von-10-3}}
\end{figure}

Figure \ref{fig:Kxy-h-Von-10-3} shows this sign jump of the phonon Hall conductivity. The role small magnetic field played in our system is to perturb our system at $\Gamma$ point to induce circular polarisation like the ``run away'' mode in the work by L\"u $\emph{et~al.}$, for the current-induced $\tilde{A}(\boldsymbol{q})$ is 0 there due to the translational symmetry. Therefore, the magnetic field determines the sign of the phonon Hall conductivity. Away from $\Gamma$ point, current-induced $\tilde{A}(\boldsymbol{q})$ starts to affect the system so that there is a discontinuity of $\kappa_{xy}$.
In section II, we said we ignore the Joule heating effect.  However, in practice, Joule heating always exists without special flowing direction.  Therefore, it will not prevent us from observing PHE. We simply prepare a sample with temperature gradient in a direction, let electric current flow parallel to this temperature gradient, and apply small magnetic field twice with opposite direction, then measure the temperature differences in the direction transverse to the current flow. The Joule heating effect does not change sign while the Hall effect changes sign.  From this, we can deduce the pure Hall contribution. 


\section{Conclusion}

In summary, we have proposed a mechanism of PHE induced by the electric
current. Compared with other PHEs, no significant magnetic field is needed in
our system.  The Chern numbers of some phonon branches are not 0, but the total Chern number of all the branches are still 0. The property of our system is that  for a suitable range of the drift velocities, the phonon Hall conductivity has a linear relation on the drift velocity, which is proportional to the applied current.

\section*{Acknowledgments}
We thanks Prof.~Jingtao L\"u  and Prof.~Lifa Zhang for discussions.  J.-S. W. is supported by a FRC grant R-144-000-402-114 and an MOE tier 2 grant
R-144-000-411-112. Z. GAO acknowledges the financial support from FRC tier 1 funding of Singapore (grant no. R-144-000-402-114).  

\appendix

\section{Dynamic matrix and EPI matrix elements}
Starting from basic coupling matrix between two atoms in x direction $K_x$, we can construct dynamic matrix of our lattice model \cite{PhysRevLett.105.225901}.
In our coordinates, unit cell lattice vectors are $\boldsymbol{a_1}=(3a/2, \sqrt{3}a/2)$ and $\boldsymbol{a_2}=(3a/2, -\sqrt{3}a/2)$. The explicit coupling matrices among three nearest pair can be obtained by a rotation matrix $U$ which are $K_{01}=U(\pi/3)K_xU(-\pi/3)$, $K_{02}=U(-\pi/3)K_xU(\pi/3)$ and $K_{03}=U(\pi)K_xU(-\pi)$ respectively. Based on these matrices, we can construct five coupling matrices between unit cells.
\begin{equation}
K_0 = \begin{pmatrix} K_{01}+K_{02}+K_{03} &-K_{03} \\ -K_{03} &K_{01}+K_{02}+K_{03}\end{pmatrix}, 
\end{equation}
\begin{equation}
K_1 = \begin{pmatrix} 0 &0 \\ -K_{02} &0 \end{pmatrix},
K_2 = \begin{pmatrix} 0 &0 \\ -K_{01} &0 \end{pmatrix},
\end{equation}
\begin{equation}
K_3 = \begin{pmatrix} 0 &-K_{02} \\ 0 &0 \end{pmatrix},
K_4 = \begin{pmatrix} 0 &-K_{01} \\ 0 &0 \end{pmatrix}.
\end{equation}
Then the dynamic matrix is
\begin{equation}
\begin{aligned} 
\bar{K}_{\boldsymbol{q}}&=K_0+K_1e^{i(3q_xa/2-\sqrt{3}q_ya/2)}+K_2e^{i(3q_xa/2+\sqrt{3}q_ya/2)}\\
&+K_3e^{-i(3q_xa/2-\sqrt{3}q_ya/2)}+K_4e^{-i(3q_xa/2+\sqrt{3}q_ya/2)}.
\end{aligned}
\end{equation}

To calculate the non-adiabatic self energy $\Pi^{\rm NA}_{\boldsymbol{q}}$, we need to know EPI matrix in reciprocal space. By transforming $\hat{H}_{\rm epi}$ into reciprocal space, we can extract tensor elements. We use A,B to represent two atoms in a unit cell and \{Ax, Ay, Bx, By\} to represent four degrees of freedom of EPI in our lattice model. Then the reciprocal EPI matrix elements are
\begin{equation}
\Xi^{\rm Ax}_{\rm AB}(\boldsymbol{k},\boldsymbol{q}) = -J_1[e^{ik_xa/2}{\rm cos}(\sqrt{3}k_ya/2)-e^{-ik_xa}],
\end{equation}
\begin{equation}
\Xi^{\rm Ay}_{\rm AB}(\boldsymbol{k},\boldsymbol{q}) = -J_1\sqrt{3}ie^{ik_xa/2}{\rm sin}(\sqrt{3}k_ya/2),
\end{equation}
\begin{equation}
\begin{aligned}
\Xi^{\rm Bx}_{\rm AB}(\boldsymbol{k},\boldsymbol{q}) = &J_1[e^{i(k_x+q_x)a/2}{\rm cos}(\sqrt{3}(k_y+q_y)a/2) \\& -e^{-i(k_x+q_x)a}],
\end{aligned}
\end{equation}
\begin{equation}
\Xi^{\rm By}_{\rm AB}(\boldsymbol{k},\boldsymbol{q}) = J_1\sqrt{3}ie^{i(k_x+q_x)a/2}{\rm sin}(\sqrt{3}(k_y+q_y)a/2),
\end{equation}
\begin{equation}
\Xi^j_{\rm BA}(\boldsymbol{k},\boldsymbol{q}) = \big(\Xi^j_{\rm AB}(\boldsymbol{k}+\boldsymbol{q},-\boldsymbol{q})\big)^*, j={\rm \{Ax, Ay, Bx, By\}}.
\end{equation}
and other elements are all zero.

\section{Equation of motion containing $A$ matrix}
For a general electron-phonon system, there is a generalized Langevin equation
describing the atoms' movement \cite{doi:10.1063/1.4917017}:
\begin{equation}
\ddot{\boldsymbol{u}}=-K\boldsymbol{u}-\int^t\Pi^r_{\rm epi}(t-t^{\prime})\boldsymbol{u}(t^{\prime})dt^{\prime} + \xi.
\end{equation}
Here we do not consider the bath contribution and set the noise term $\xi$ to
zero, for our system is infinitely large. We can define
$d\Gamma(t)/dt\equiv\Pi^r_{\rm epi}(t)$ and integrate by parts so that the
equation of motion becomes:
\begin{equation}
\ddot{\boldsymbol{u}}=-K\boldsymbol{u}-\int^t\Gamma(t-t^{\prime})\dot{\boldsymbol{u}}(t^{\prime})dt^{\prime}.
\end{equation}
Next we apply a Markov approximation to $\Gamma(t-t^{\prime})$ so that
$\Gamma(t-t^{\prime})\approx 4A(t^{\prime})\delta(t-t^{\prime})$ (factor 4 is
for consistency).  The final expression of the equation of motion will be:
\begin{equation}
\ddot{\boldsymbol{u}}=-K\boldsymbol{u}-2A\dot{\boldsymbol{u}},
\end{equation}
which is used in section II.

\section{Berry curvature}
Usually there are two ways of calculating the Berry curvature, one is the
explicit way by inserting the completeness identity into the definition of the
Berry curvature. In our system,  the explicit formula is 
\begin{equation}
\Omega_{i}=-\text{Im}\sum\limits_{i^{\prime}\ne i}\frac{\bar{\psi}_{i}\frac{\partial H_{\rm eff}}{\partial q_x}\psi_{i^{\prime}}\bar{\psi}_{i^{\prime}}\frac{\partial H_{\rm eff}}{\partial q_y}\psi_{i}-(q_x\leftrightarrow q_y)}{(\omega_{i}-\omega_{i^{\prime}})^2}.
\end{equation} 
However, to calculate the partial derivative of $H_{\rm eff}$, we need
numerical differentiation which will cost a large amount of computation to be
precise enough. Therefore we choose another way, a geometric way by dividing
the Brillouin zone into plaquettes each consisting of four points on a square
with area $\Delta S$ and calculating the Berry phase around them
\cite{fukui2005chern,vanderbilt2018berry}.
\begin{equation}
\phi=-\text{Im}\ln(\bar{\psi}_1\psi_2\bar{\psi}_2\psi_3\bar{\psi}_3\psi_4\bar{\psi}_4\psi_1),
\end{equation} 
Compared with the Hermitian case, we have replaced the Hermitian conjugate of the
eigenvector by the left eigenvector. If investigated further, we find that this
replacement is not correct for $\bar{\psi}_1\psi_2\ne(\bar{\psi}_2\psi_1)^*$.
This break of the equality, a fundamental property of the inner product in
Hilbert space, will invalidate Stokes' theorem so that we cannot obtain
Berry curvature through Berry phase. To overcome this, we define a new version
of inner product:
\begin{equation}
\langle \bar{\psi}_1\psi_2\rangle \equiv\frac{\bar{\psi}_1\psi_2+(\bar{\psi}_2\psi_1)^*}{2}.
\end{equation}
With this definition, property of inner product in Hilbert space and validity of Stokes' theorem are restored. Then the Berry curvature is calculated by:
\begin{equation}
\Omega=\lim\limits_{\Delta S\rightarrow0}\frac{-\text{Im}\ln\bigl(\langle \bar{\psi}_1\psi_2\rangle \langle \bar{\psi}_2\psi_3\rangle \langle \bar{\psi}_3\psi_4\rangle \langle \bar{\psi}_4\psi_1\rangle \bigr)}{\Delta S}.
\end{equation}
One can show that the two ways computing the Berry curvature are mathematically equivalent. 
\vfill  

\bibliography{references}

\begin{thebibliography}{23}%
\makeatletter
\providecommand \@ifxundefined [1]{%
 \@ifx{#1\undefined}
}%
\providecommand \@ifnum [1]{%
 \ifnum #1\expandafter \@firstoftwo
 \else \expandafter \@secondoftwo
 \fi
}%
\providecommand \@ifx [1]{%
 \ifx #1\expandafter \@firstoftwo
 \else \expandafter \@secondoftwo
 \fi
}%
\providecommand \natexlab [1]{#1}%
\providecommand \enquote  [1]{``#1''}%
\providecommand \bibnamefont  [1]{#1}%
\providecommand \bibfnamefont [1]{#1}%
\providecommand \citenamefont [1]{#1}%
\providecommand \href@noop [0]{\@secondoftwo}%
\providecommand \href [0]{\begingroup \@sanitize@url \@href}%
\providecommand \@href[1]{\@@startlink{#1}\@@href}%
\providecommand \@@href[1]{\endgroup#1\@@endlink}%
\providecommand \@sanitize@url [0]{\catcode `\\12\catcode `\$12\catcode
  `\&12\catcode `\#12\catcode `\^12\catcode `\_12\catcode `\%12\relax}%
\providecommand \@@startlink[1]{}%
\providecommand \@@endlink[0]{}%
\providecommand \url  [0]{\begingroup\@sanitize@url \@url }%
\providecommand \@url [1]{\endgroup\@href {#1}{\urlprefix }}%
\providecommand \urlprefix  [0]{URL }%
\providecommand \Eprint [0]{\href }%
\providecommand \doibase [0]{https://doi.org/}%
\providecommand \selectlanguage [0]{\@gobble}%
\providecommand \bibinfo  [0]{\@secondoftwo}%
\providecommand \bibfield  [0]{\@secondoftwo}%
\providecommand \translation [1]{[#1]}%
\providecommand \BibitemOpen [0]{}%
\providecommand \bibitemStop [0]{}%
\providecommand \bibitemNoStop [0]{.\EOS\space}%
\providecommand \EOS [0]{\spacefactor3000\relax}%
\providecommand \BibitemShut  [1]{\csname bibitem#1\endcsname}%
\let\auto@bib@innerbib\@empty
\bibitem [{\citenamefont {Strohm}\ \emph {et~al.}(2005)\citenamefont {Strohm},
  \citenamefont {Rikken},\ and\ \citenamefont
  {Wyder}}]{strohm2005phenomenological}%
  \BibitemOpen
  \bibfield  {author} {\bibinfo {author} {\bibfnamefont {C.}~\bibnamefont
  {Strohm}}, \bibinfo {author} {\bibfnamefont {G.}~\bibnamefont {Rikken}},\
  and\ \bibinfo {author} {\bibfnamefont {P.}~\bibnamefont {Wyder}},\ }\bibfield
   {title} {\bibinfo {title} {Phenomenological evidence for the phonon hall
  effect},\ }\href@noop {} {\bibfield  {journal} {\bibinfo  {journal} {Physical
  review letters}\ }\textbf {\bibinfo {volume} {95}},\ \bibinfo {pages}
  {155901} (\bibinfo {year} {2005})}\BibitemShut {NoStop}%
\bibitem [{\citenamefont {Zhang}\ \emph {et~al.}(2010)\citenamefont {Zhang},
  \citenamefont {Ren}, \citenamefont {Wang},\ and\ \citenamefont
  {Li}}]{PhysRevLett.105.225901}%
  \BibitemOpen
  \bibfield  {author} {\bibinfo {author} {\bibfnamefont {L.}~\bibnamefont
  {Zhang}}, \bibinfo {author} {\bibfnamefont {J.}~\bibnamefont {Ren}}, \bibinfo
  {author} {\bibfnamefont {J.-S.}\ \bibnamefont {Wang}},\ and\ \bibinfo
  {author} {\bibfnamefont {B.}~\bibnamefont {Li}},\ }\bibfield  {title}
  {\bibinfo {title} {Topological nature of the phonon hall effect},\ }\href
  {https://doi.org/10.1103/PhysRevLett.105.225901} {\bibfield  {journal}
  {\bibinfo  {journal} {Phys. Rev. Lett.}\ }\textbf {\bibinfo {volume} {105}},\
  \bibinfo {pages} {225901} (\bibinfo {year} {2010})}\BibitemShut {NoStop}%
\bibitem [{\citenamefont {Qin}\ \emph {et~al.}(2012)\citenamefont {Qin},
  \citenamefont {Zhou},\ and\ \citenamefont {Shi}}]{PhysRevB.86.104305}%
  \BibitemOpen
  \bibfield  {author} {\bibinfo {author} {\bibfnamefont {T.}~\bibnamefont
  {Qin}}, \bibinfo {author} {\bibfnamefont {J.}~\bibnamefont {Zhou}},\ and\
  \bibinfo {author} {\bibfnamefont {J.}~\bibnamefont {Shi}},\ }\bibfield
  {title} {\bibinfo {title} {Berry curvature and the phonon hall effect},\
  }\href {https://doi.org/10.1103/PhysRevB.86.104305} {\bibfield  {journal}
  {\bibinfo  {journal} {Phys. Rev. B}\ }\textbf {\bibinfo {volume} {86}},\
  \bibinfo {pages} {104305} (\bibinfo {year} {2012})}\BibitemShut {NoStop}%
\bibitem [{\citenamefont {Saito}\ \emph {et~al.}(2019)\citenamefont {Saito},
  \citenamefont {Misaki}, \citenamefont {Ishizuka},\ and\ \citenamefont
  {Nagaosa}}]{saito2019berry}%
  \BibitemOpen
  \bibfield  {author} {\bibinfo {author} {\bibfnamefont {T.}~\bibnamefont
  {Saito}}, \bibinfo {author} {\bibfnamefont {K.}~\bibnamefont {Misaki}},
  \bibinfo {author} {\bibfnamefont {H.}~\bibnamefont {Ishizuka}},\ and\
  \bibinfo {author} {\bibfnamefont {N.}~\bibnamefont {Nagaosa}},\ }\bibfield
  {title} {\bibinfo {title} {Berry phase of phonons and thermal hall effect in
  nonmagnetic insulators},\ }\href@noop {} {\bibfield  {journal} {\bibinfo
  {journal} {Physical Review Letters}\ }\textbf {\bibinfo {volume} {123}},\
  \bibinfo {pages} {255901} (\bibinfo {year} {2019})}\BibitemShut {NoStop}%
\bibitem [{\citenamefont {Li}\ \emph {et~al.}(2020)\citenamefont {Li},
  \citenamefont {Fauqu\'e}, \citenamefont {Zhu},\ and\ \citenamefont
  {Behnia}}]{PhysRevLett.124.105901}%
  \BibitemOpen
  \bibfield  {author} {\bibinfo {author} {\bibfnamefont {X.}~\bibnamefont
  {Li}}, \bibinfo {author} {\bibfnamefont {B.}~\bibnamefont {Fauqu\'e}},
  \bibinfo {author} {\bibfnamefont {Z.}~\bibnamefont {Zhu}},\ and\ \bibinfo
  {author} {\bibfnamefont {K.}~\bibnamefont {Behnia}},\ }\bibfield  {title}
  {\bibinfo {title} {Phonon thermal hall effect in strontium titanate},\ }\href
  {https://doi.org/10.1103/PhysRevLett.124.105901} {\bibfield  {journal}
  {\bibinfo  {journal} {Phys. Rev. Lett.}\ }\textbf {\bibinfo {volume} {124}},\
  \bibinfo {pages} {105901} (\bibinfo {year} {2020})}\BibitemShut {NoStop}%
\bibitem [{\citenamefont {Zhang}\ \emph {et~al.}(2019)\citenamefont {Zhang},
  \citenamefont {Zhang}, \citenamefont {Okamoto},\ and\ \citenamefont
  {Xiao}}]{PhysRevLett.123.167202}%
  \BibitemOpen
  \bibfield  {author} {\bibinfo {author} {\bibfnamefont {X.}~\bibnamefont
  {Zhang}}, \bibinfo {author} {\bibfnamefont {Y.}~\bibnamefont {Zhang}},
  \bibinfo {author} {\bibfnamefont {S.}~\bibnamefont {Okamoto}},\ and\ \bibinfo
  {author} {\bibfnamefont {D.}~\bibnamefont {Xiao}},\ }\bibfield  {title}
  {\bibinfo {title} {Thermal hall effect induced by magnon-phonon
  interactions},\ }\href {https://doi.org/10.1103/PhysRevLett.123.167202}
  {\bibfield  {journal} {\bibinfo  {journal} {Phys. Rev. Lett.}\ }\textbf
  {\bibinfo {volume} {123}},\ \bibinfo {pages} {167202} (\bibinfo {year}
  {2019})}\BibitemShut {NoStop}%
\bibitem [{\citenamefont {Agarwalla}\ \emph {et~al.}(2011)\citenamefont
  {Agarwalla}, \citenamefont {Zhang}, \citenamefont {Wang},\ and\ \citenamefont
  {Li}}]{agarwalla2011phonon}%
  \BibitemOpen
  \bibfield  {author} {\bibinfo {author} {\bibfnamefont {B.~K.}\ \bibnamefont
  {Agarwalla}}, \bibinfo {author} {\bibfnamefont {L.}~\bibnamefont {Zhang}},
  \bibinfo {author} {\bibfnamefont {J.-S.}\ \bibnamefont {Wang}},\ and\
  \bibinfo {author} {\bibfnamefont {B.}~\bibnamefont {Li}},\ }\bibfield
  {title} {\bibinfo {title} {Phonon hall effect in ionic crystals in the
  presence of static magnetic field},\ }\href@noop {} {\bibfield  {journal}
  {\bibinfo  {journal} {The European Physical Journal B}\ }\textbf {\bibinfo
  {volume} {81}},\ \bibinfo {pages} {197} (\bibinfo {year} {2011})}\BibitemShut
  {NoStop}%
\bibitem [{\citenamefont {Chen}\ \emph {et~al.}(2020)\citenamefont {Chen},
  \citenamefont {Kivelson},\ and\ \citenamefont {Sun}}]{chen2020enhanced}%
  \BibitemOpen
  \bibfield  {author} {\bibinfo {author} {\bibfnamefont {J.-Y.}\ \bibnamefont
  {Chen}}, \bibinfo {author} {\bibfnamefont {S.~A.}\ \bibnamefont {Kivelson}},\
  and\ \bibinfo {author} {\bibfnamefont {X.-Q.}\ \bibnamefont {Sun}},\
  }\bibfield  {title} {\bibinfo {title} {Enhanced thermal hall effect in nearly
  ferroelectric insulators},\ }\href@noop {} {\bibfield  {journal} {\bibinfo
  {journal} {Physical Review Letters}\ }\textbf {\bibinfo {volume} {124}},\
  \bibinfo {pages} {167601} (\bibinfo {year} {2020})}\BibitemShut {NoStop}%
\bibitem [{\citenamefont {Lü}\ \emph {et~al.}(2010)\citenamefont {Lü},
  \citenamefont {Brandbyge},\ and\ \citenamefont
  {Hedegård}}]{doi:10.1021/nl904233u}%
  \BibitemOpen
  \bibfield  {author} {\bibinfo {author} {\bibfnamefont {J.-T.}\ \bibnamefont
  {Lü}}, \bibinfo {author} {\bibfnamefont {M.}~\bibnamefont {Brandbyge}},\
  and\ \bibinfo {author} {\bibfnamefont {P.}~\bibnamefont {Hedegård}},\
  }\bibfield  {title} {\bibinfo {title} {Blowing the fuse: Berry’s phase and
  runaway vibrations in molecular conductors},\ }\href
  {https://doi.org/10.1021/nl904233u} {\bibfield  {journal} {\bibinfo
  {journal} {Nano Letters}\ }\textbf {\bibinfo {volume} {10}},\ \bibinfo
  {pages} {1657} (\bibinfo {year} {2010})},\ \Eprint
  {https://arxiv.org/abs/https://doi.org/10.1021/nl904233u}
  {https://doi.org/10.1021/nl904233u} \BibitemShut {NoStop}%
\bibitem [{\citenamefont {Wang}\ and\ \citenamefont
  {Zhang}(2009)}]{PhysRevB.80.012301}%
  \BibitemOpen
  \bibfield  {author} {\bibinfo {author} {\bibfnamefont {J.-S.}\ \bibnamefont
  {Wang}}\ and\ \bibinfo {author} {\bibfnamefont {L.}~\bibnamefont {Zhang}},\
  }\bibfield  {title} {\bibinfo {title} {Phonon hall thermal conductivity from
  the green-kubo formula},\ }\href {https://doi.org/10.1103/PhysRevB.80.012301}
  {\bibfield  {journal} {\bibinfo  {journal} {Phys. Rev. B}\ }\textbf {\bibinfo
  {volume} {80}},\ \bibinfo {pages} {012301} (\bibinfo {year}
  {2009})}\BibitemShut {NoStop}%
\bibitem [{\citenamefont {Sheng}\ \emph {et~al.}(2006)\citenamefont {Sheng},
  \citenamefont {Sheng},\ and\ \citenamefont {Ting}}]{sheng2006theory}%
  \BibitemOpen
  \bibfield  {author} {\bibinfo {author} {\bibfnamefont {L.}~\bibnamefont
  {Sheng}}, \bibinfo {author} {\bibfnamefont {D.}~\bibnamefont {Sheng}},\ and\
  \bibinfo {author} {\bibfnamefont {C.}~\bibnamefont {Ting}},\ }\bibfield
  {title} {\bibinfo {title} {Theory of the phonon hall effect in paramagnetic
  dielectrics},\ }\href@noop {} {\bibfield  {journal} {\bibinfo  {journal}
  {Physical review letters}\ }\textbf {\bibinfo {volume} {96}},\ \bibinfo
  {pages} {155901} (\bibinfo {year} {2006})}\BibitemShut {NoStop}%
\bibitem [{\citenamefont {Kagan}\ and\ \citenamefont
  {Maksimov}(2008)}]{kagan2008anomalous}%
  \BibitemOpen
  \bibfield  {author} {\bibinfo {author} {\bibfnamefont {Y.}~\bibnamefont
  {Kagan}}\ and\ \bibinfo {author} {\bibfnamefont {L.}~\bibnamefont
  {Maksimov}},\ }\bibfield  {title} {\bibinfo {title} {Anomalous hall effect
  for the phonon heat conductivity in paramagnetic dielectrics},\ }\href@noop
  {} {\bibfield  {journal} {\bibinfo  {journal} {Physical review letters}\
  }\textbf {\bibinfo {volume} {100}},\ \bibinfo {pages} {145902} (\bibinfo
  {year} {2008})}\BibitemShut {NoStop}%
\bibitem [{\citenamefont {Mead}(1992)}]{mead1992geometric}%
  \BibitemOpen
  \bibfield  {author} {\bibinfo {author} {\bibfnamefont {C.~A.}\ \bibnamefont
  {Mead}},\ }\bibfield  {title} {\bibinfo {title} {The geometric phase in
  molecular systems},\ }\href@noop {} {\bibfield  {journal} {\bibinfo
  {journal} {Reviews of modern physics}\ }\textbf {\bibinfo {volume} {64}},\
  \bibinfo {pages} {51} (\bibinfo {year} {1992})}\BibitemShut {NoStop}%
\bibitem [{\citenamefont {Mead}\ and\ \citenamefont
  {Truhlar}(1979)}]{doi:10.1063/1.437734}%
  \BibitemOpen
  \bibfield  {author} {\bibinfo {author} {\bibfnamefont {C.~A.}\ \bibnamefont
  {Mead}}\ and\ \bibinfo {author} {\bibfnamefont {D.~G.}\ \bibnamefont
  {Truhlar}},\ }\bibfield  {title} {\bibinfo {title} {On the determination of
  born–oppenheimer nuclear motion wave functions including complications due
  to conical intersections and identical nuclei},\ }\href
  {https://doi.org/10.1063/1.437734} {\bibfield  {journal} {\bibinfo  {journal}
  {The Journal of Chemical Physics}\ }\textbf {\bibinfo {volume} {70}},\
  \bibinfo {pages} {2284} (\bibinfo {year} {1979})},\ \Eprint
  {https://arxiv.org/abs/https://doi.org/10.1063/1.437734}
  {https://doi.org/10.1063/1.437734} \BibitemShut {NoStop}%
\bibitem [{\citenamefont {Zhang}(2016)}]{zhang2016berry}%
  \BibitemOpen
  \bibfield  {author} {\bibinfo {author} {\bibfnamefont {L.}~\bibnamefont
  {Zhang}},\ }\bibfield  {title} {\bibinfo {title} {Berry curvature and various
  thermal hall effects},\ }\href@noop {} {\bibfield  {journal} {\bibinfo
  {journal} {New Journal of Physics}\ }\textbf {\bibinfo {volume} {18}},\
  \bibinfo {pages} {103039} (\bibinfo {year} {2016})}\BibitemShut {NoStop}%
\bibitem [{\citenamefont {Giustino}(2017)}]{RevModPhys.89.015003}%
  \BibitemOpen
  \bibfield  {author} {\bibinfo {author} {\bibfnamefont {F.}~\bibnamefont
  {Giustino}},\ }\bibfield  {title} {\bibinfo {title} {Electron-phonon
  interactions from first principles},\ }\href
  {https://doi.org/10.1103/RevModPhys.89.015003} {\bibfield  {journal}
  {\bibinfo  {journal} {Rev. Mod. Phys.}\ }\textbf {\bibinfo {volume} {89}},\
  \bibinfo {pages} {015003} (\bibinfo {year} {2017})}\BibitemShut {NoStop}%
\bibitem [{\citenamefont {Lü}\ \emph {et~al.}(2015)\citenamefont {Lü},
  \citenamefont {Zhou}, \citenamefont {Jiang},\ and\ \citenamefont
  {Wang}}]{doi:10.1063/1.4917017}%
  \BibitemOpen
  \bibfield  {author} {\bibinfo {author} {\bibfnamefont {J.-T.}\ \bibnamefont
  {Lü}}, \bibinfo {author} {\bibfnamefont {H.}~\bibnamefont {Zhou}}, \bibinfo
  {author} {\bibfnamefont {J.-W.}\ \bibnamefont {Jiang}},\ and\ \bibinfo
  {author} {\bibfnamefont {J.-S.}\ \bibnamefont {Wang}},\ }\bibfield  {title}
  {\bibinfo {title} {Effects of electron-phonon interaction on thermal and
  electrical transport through molecular nano-conductors},\ }\href
  {https://doi.org/10.1063/1.4917017} {\bibfield  {journal} {\bibinfo
  {journal} {AIP Advances}\ }\textbf {\bibinfo {volume} {5}},\ \bibinfo {pages}
  {053204} (\bibinfo {year} {2015})},\ \Eprint
  {https://arxiv.org/abs/https://doi.org/10.1063/1.4917017}
  {https://doi.org/10.1063/1.4917017} \BibitemShut {NoStop}%
\bibitem [{\citenamefont {Jiang}\ and\ \citenamefont
  {Wang}(2011)}]{doi:10.1063/1.3671069}%
  \BibitemOpen
  \bibfield  {author} {\bibinfo {author} {\bibfnamefont {J.-W.}\ \bibnamefont
  {Jiang}}\ and\ \bibinfo {author} {\bibfnamefont {J.-S.}\ \bibnamefont
  {Wang}},\ }\bibfield  {title} {\bibinfo {title} {Joule heating and
  thermoelectric properties in short single-walled carbon nanotubes:
  Electron-phonon interaction effect},\ }\href
  {https://doi.org/10.1063/1.3671069} {\bibfield  {journal} {\bibinfo
  {journal} {Journal of Applied Physics}\ }\textbf {\bibinfo {volume} {110}},\
  \bibinfo {pages} {124319} (\bibinfo {year} {2011})},\ \Eprint
  {https://arxiv.org/abs/https://doi.org/10.1063/1.3671069}
  {https://doi.org/10.1063/1.3671069} \BibitemShut {NoStop}%
\bibitem [{\citenamefont {Ziman}(2001)}]{ziman2001electrons}%
  \BibitemOpen
  \bibfield  {author} {\bibinfo {author} {\bibfnamefont {J.~M.}\ \bibnamefont
  {Ziman}},\ }\href@noop {} {\emph {\bibinfo {title} {Electrons and phonons:
  the theory of transport phenomena in solids}}}\ (\bibinfo  {publisher}
  {Oxford university press},\ \bibinfo {year} {2001})\BibitemShut {NoStop}%
\bibitem [{\citenamefont {Peng}\ and\ \citenamefont
  {Wang}(2018)}]{peng2018current}%
  \BibitemOpen
  \bibfield  {author} {\bibinfo {author} {\bibfnamefont {J.}~\bibnamefont
  {Peng}}\ and\ \bibinfo {author} {\bibfnamefont {J.-S.}\ \bibnamefont
  {Wang}},\ }\bibfield  {title} {\bibinfo {title} {Current-induced heat
  transfer in double-layer graphene},\ }\href@noop {} {\bibfield  {journal}
  {\bibinfo  {journal} {arXiv preprint arXiv:1805.09493}\ } (\bibinfo {year}
  {2018})}\BibitemShut {NoStop}%
\bibitem [{\citenamefont {Liu}\ \emph {et~al.}(2017)\citenamefont {Liu},
  \citenamefont {Xu}, \citenamefont {Zhang},\ and\ \citenamefont
  {Duan}}]{liu2017model}%
  \BibitemOpen
  \bibfield  {author} {\bibinfo {author} {\bibfnamefont {Y.}~\bibnamefont
  {Liu}}, \bibinfo {author} {\bibfnamefont {Y.}~\bibnamefont {Xu}}, \bibinfo
  {author} {\bibfnamefont {S.-C.}\ \bibnamefont {Zhang}},\ and\ \bibinfo
  {author} {\bibfnamefont {W.}~\bibnamefont {Duan}},\ }\bibfield  {title}
  {\bibinfo {title} {Model for topological phononics and phonon diode},\
  }\href@noop {} {\bibfield  {journal} {\bibinfo  {journal} {Physical Review
  B}\ }\textbf {\bibinfo {volume} {96}},\ \bibinfo {pages} {064106} (\bibinfo
  {year} {2017})}\BibitemShut {NoStop}%
\bibitem [{\citenamefont {Fukui}\ \emph {et~al.}(2005)\citenamefont {Fukui},
  \citenamefont {Hatsugai},\ and\ \citenamefont {Suzuki}}]{fukui2005chern}%
  \BibitemOpen
  \bibfield  {author} {\bibinfo {author} {\bibfnamefont {T.}~\bibnamefont
  {Fukui}}, \bibinfo {author} {\bibfnamefont {Y.}~\bibnamefont {Hatsugai}},\
  and\ \bibinfo {author} {\bibfnamefont {H.}~\bibnamefont {Suzuki}},\
  }\bibfield  {title} {\bibinfo {title} {Chern numbers in discretized brillouin
  zone: efficient method of computing (spin) hall conductances},\ }\href@noop
  {} {\bibfield  {journal} {\bibinfo  {journal} {Journal of the Physical
  Society of Japan}\ }\textbf {\bibinfo {volume} {74}},\ \bibinfo {pages}
  {1674} (\bibinfo {year} {2005})}\BibitemShut {NoStop}%
\bibitem [{\citenamefont {Vanderbilt}(2018)}]{vanderbilt2018berry}%
  \BibitemOpen
  \bibfield  {author} {\bibinfo {author} {\bibfnamefont {D.}~\bibnamefont
  {Vanderbilt}},\ }\href@noop {} {\emph {\bibinfo {title} {Berry Phases in
  Electronic Structure Theory: Electric Polarization, Orbital Magnetization and
  Topological Insulators}}}\ (\bibinfo  {publisher} {Cambridge University
  Press},\ \bibinfo {year} {2018})\BibitemShut {NoStop}%
\end{thebibliography}%

\end{document}